# *Agave villalobosii* (SECCIÓN *DITEPALAE*, AGAVACEAE), UNA ESPECIE NUEVA DE LA MESETA CENTRAL MEXICANA DE AGUASCALIENTES Y ZACATECAS

Gutiérrez-Gutiérrez, B. G.[1], Luquin-Arce, J. L.[2], Padilla-Lepe, J.[1], Nieves Hernández, G.[1], Oropeza-Gutiérrez, P. J. y Vázquez-García, J. A.[1], [1]Herbario IBUG, Instituto de Botánica, Departamento de Botánica y Zoología, Centro Universitario de Ciencias Biológicas y Agropecuarias, Universidad de Guadalajara, km 15.5 Guadalajara-Nogales, Las Agujas, Nextipac, Zapopan, C.P. 45200, Jalisco, México. [2]Investigador independiente, Rancho Agroecológico Chihuila, El Cabezón, Ameca, Jalisco. Autor para la correspondencia jantonio.vazquez@academicos.udg.mx

**RESUMEN**

Se describe e ilustra *Agave villalobosii* sp. nov. (sección *Ditepalae*, Agavaceae, Asparagales) de la meseta central mexicana en Aguascalientes y sur de Zacatecas, México. Se asemeja a *A. flexispina* en cuanto a color y apariencia general de las rosetas. Sin embargo, se diferencia de esta última por tener menos hojas con dientes más espaciados, panículas más compactas y cortas, ramas laterales más inclinadas con respecto del plano horizontal| y cápsulas subglobosas a ampliamente elipsoides. Se proporciona un mapa de distribución para ambas especies. La especie se diagnosticó preliminarmente como en peligro crítico.

**ABSTRACT**

*Agave villalobosii sp. nov.* (sect. *Ditepalae*, Agavaceae, Asparagales) from the Mexican central plain in Aguascalientes and southern Zacatecas, Mexico, is described and illustrated. It resembles *A. flexispina* in terms of color and the general appearance of its rosettes. However, it differs from the latter in having fewer leaves with more widely spaced teeth, more compact and shorter panicles with more inclined lateral branches with respect to the horizontal plane, and subglobose to broadly ellipsoid capsules. A distribution map for both species is provided. The species was preliminary assessed as critically endangered.

## INTRODUCCIÓN

*Agave* L. en México, con cerca de 170 especies, se encuentra entre los diez géneros más diversos de la flora mexicana y más del 80% de las especies son endémicas (Gentry 1982, Vázquez-García *et al*. 2016; García-Mendoza *et al*. 2019, Thiede *et al*. 2020).

*Agave* sección *Ditepalae* Hochstätter (subgen. *Agave*) incluye 16 especies (incluyendo la propuesta aquí como nueva): *A. applanata* Trel., *A. chysantha* Peebles, *A. colorata* Gentry, *A. delamateri* W. C. Hodgson & Slauson, *A. durangensis* Gentry, *A. flexispina* Trel., *A. fortiflora* Gentry, *A. lyobaa* García-Mendoza & S. Franco, *A. murpheyi* F. Gibson, *A. palmeri* Engelmann, *A. temacapulinensis* A.Vázquez & Cházaro, *A. shrevei* Gentry, *A. verdensis* W. C. Hodgson & Salywon, *A. wocomahi* Gentry y *A. yavapaiensis* W. C. Hodgson & Salywon (Gentry 1982; McVaugh 1989; Hernández-Vera *et al*. 2007; González-Elizondo 2009; Vázquez-García *et al*. 2012; Hodgson y Slauson 1995; Hodgson y Salywon 2013; García-Mendoza *et al*. 2019). Seis especies de esta sección ocurren en Estados Unidos, 11 en México y, de estos, cinco en el occidente de México (Gentry 1982, Vázquez-García *et al*. 2007a, Thiede 2020). Esta sección no es monofilética, ya que mientras *A. colorata* pertenece al clado denominado Grupo IV, *A. delamateri* se ubica en un clado distinto denominado Grupo V (Jiménez-Barrón *et al*. 2020). Se requiere incluir más especies de esta sección en estudios moleculares para entender las relaciones intraseccionales entre especies.

Las especies de *Agave* sección *Ditepalae* se diferencian de otros grupos de agaves por ser plantas pequeñas a grandes con rosetas mayormente glaucas, claras a blanquecinas, con polinización cruzada, únicas o escasamente surculosas; hojas siempre bien armadas, firmes a rígidas; panículas generalmente abiertas, con brácteas reflexas y escariosas; flores generalmente rojizas en botón y amarillas en la antesis, con tubos florales profundos; tépalos del mismo tamaño que el tubo o más cortos, dimórficos, coriáceos al secarse, persistentes y erectos, el más grande superpuesto al interno (Gentry 1982). Este grupo distingue y caracteriza el elemento suculento en la flora de la Sierra Madre Occidental, algunos con distribución atípica (Gentry 1982). Su distribución se aprecia desde los valles desérticos del sur de Estados Unidos, hacia el sur por la vertiente del Océano Pacífico por los estados de Durango, Sinaloa, y hacia el centro del país por Zacatecas, Aguascalientes, Jalisco, Querétaro, Hidalgo y Puebla.

Para el estado de Zacatecas se reportan cuatro especies de *Agave* de la sección *Ditepalae*: *A. applanata*, *A. durangensis*, *A. flexispina* y *A. woccomahi*. Sin embargo, para el estado de Aguascalientes no se ha reportado ninguna especie de esta sección (Gentry 1982, De La Cerda-Lemus 2004, 2007; Vázquez-García *et al*. 2007b; Villaseñor 2016). Un examen preliminar de poblaciones de *Agave* spp. de los estados de Aguascalientes y Zacatecas permite concluir que una población de *Agave* sección *Ditepalae* no corresponde con ninguna de las especies reportadas para Aguascalientes y se registra por primera vez para la entidad, mientras que tres poblaciones, de la misma sección, dos de Aguascalientes (Tepezalá y Tepetates) y una de Zacatecas (Guadalupe), constituyen un taxón aún no descrito, el cual es el objeto central de este estudio.

La finalidad de este estudio es determinar la identidad taxonómica de poblaciones de *Agave* sp. de Tepezalá, Aguascalientes y Guadalupe, Zacatecas, y contrastar con otras entidades taxonómicas afines. Se hipotetiza que se trata de una especie nueva para la ciencia, distinta de dos especies similares como *A. applanata* y *A. flexispina*. Esta especie fue trasplantada en 2011 y cultivada en el Jardín Comunitario de la localidad El Valle de Las Delicias, Rincón de Romos, aledaño a las poblaciones *in situ* de Tepezalá y Tepetate en Aguascalientes, con la finalidad de observar plasticidad fenotípica en suelo más rico y más húmedo, habiéndose observado que las rosetas mantienen un tamaño pequeño y la inflorescencia alcanzó 1.5 m de altura mucho menor de lo observado en el campo, lo que apoya la hipótesis de que las diferencias de estas poblaciones se debe a una diferenciación genética y no a los factores abióticos, los cuales parecen no tener relación con la morfología.

**MÉTODOS**
Se realizaron dos expediciones (recolectas) botánicas: 1) En marzo de 2025 a las Llanuras de Tepezalá, y cerca de la presa El Tepetate, Aguascalientes, y 2) En octubre de 2025 a Ojuelos, Jalisco. Se estudiaron especímenes de los herbarios DES, IBUG, MEXU, NY y US. Se consultaron diversas plataformas virtuales: Biology Heritage Library: https://www.biodiversitylibrary.org, particularmente protologos de las especies más relacionadas; Global Plants Jstor: https://plants.jstor.org; IPNI (https://www.ipni.org), imágenes de alta resolución de holótipo e isótipos de *Agave flexispina* y neótipo de *A. applanata*; Naturalista (https://www.naturalista.mx), observaciones complementarias compartidas por ciudadanos; POWO (https://powo.science.kew.org), recursos confiables sobre plantas vasculares; SEINET (https://swbiodiversity.org/seinet), colecciones de especímenes de distintos herbarios; Smithsonian Botany Collections-NMNH (https://collections.nmnh.si.edu) proporcionó entendimiento de variabilidad de futos e inflorescencia de *A. flexispina*; The Agavaceae Database: (https://www.agavaceae.com), imágenes de color de flores y habitat natural *de A. flexispina*; y Tropicos (https://www.tropicos.org), información taxonómica, nomenclatural fiable sobre especies y especímenes. Se obtuvo permiso de los autores de para acceder a algunas de las fotografías usadas. Se elaboró un cuadro comparativo con dos especies morfológica y geográficamente cercanas (*A. flexispina* y *A. aplanatta* Cuadro 1.)

Se utilizó el concepto morfológico de especie (Cronquist 1978). La descripción morfológica siguió la terminología de Thiede (2020), así como el concepto adoptado para la familia Agavaceae. El mapa de distribución (Fig. 1) se elaboró con Google Earth utilizando nuestros registros de campo, las localidades de los especímenes de herbario y observaciones seleccionadas de iNaturalist (Quirino 2020, González Gallina 2025). La extensión de ocupación (AOO) y la extensión de distribución (EOO) se estimaron con un tamaño de celda de cuadrícula de 2 km² en GeoCAT (Bachman *et al*. 2011).

**RESULTADOS**
**Tratamiento Taxonómico**
*Agave villalobosii* A. Vázquez & B. Gut. sp. nov. (Figs. 1-5)

TIPO: MÉXICO: Aguascalientes: Tepezalá, Cerro El Capulín, Llanuras con pastizales, en suelos calcáreos, matorral xerófilo con *Tymophylla setifolia, Russelia* sp. y *Stenocactus* sp., *Yucca filifera*, 8 Marzo 2025 (fruto), *J. Antonio Vázquez García, Iván Villalobos Juárez, Paola J. Oropeza Gutiérrez, J. Luis Luquin Arce* y *Fábio Trigo Raya 10399* (Holotipo: IBUG; isotipo: UAA).

**Diagnosis**: *Agave villalobosii* shares with *Agave flexispina* a similar rosette color and appearance, however it differs from the later in having less numerous leaves 20–25 (–30) vs. 35–40 and these with more spaced teeth 2.0–2.2 vs. 0.7–1.5 cm apart, more compact panicles with shorter branches 3.7–6.9 × 0.5–1.1 vs. 9.2–10.6 x 0.8–0.9 cm, lateral branches less inclined with respect to the horizontal plane (x-axis) 55–65° vs. 45–50°, capsules subglobose to broadly ellipsoid and shorter (2.3–2.5 cm long) vs. oblongoid to ellipsoid and longer (3.5–4.5) cm long (Cuadro 1).

**Plantas** perennes, acaulescentes, solitarias, monocárpicas; rosetas de 50.0–70.0 cm de diámetro, altura de 38.0–40.0 cm, compactas, deltoides, rígidas de coloración verde glauco, con 12–30 hojas por roseta. **Hojas** 19.5–22.3 × 15.5–16.1 cm, lanceoladas, rígidas, fibrosas, cerosas, grisáceas verdosas; márgenes dentados cerosos; vaina 6.2–8.8 × 4.4–4.6 cm; 8–10 dientes de 4.0–8.0 × 3.0–4.0 mm, flexos, curvos hacia abajo, 1.5–2.2 cm de separación, gris rojizos, cerosos, dientes intersticiales presentes, la base de la hoja (vaina) casi tan ancha como la parte más ancha de la hoja, más angosta cerca de la vaina; el ápice acuminado; espinas 3.4–6.1 × 0.3–0.5 cm, aciculares, onduladas, grisáceas. **Inflorescencias** en panícula de 1.50–1.65 × 0.20–0.23 m, porción infértil 1.20 m; porción fértil 0.32 m; panículas compactas, firmes; brácteas pedunculares 4.0 × 13.0 cm, triangulares, poco persistentes cuando secan; raquis 60.0–70.0 × 2.0–3.0 cm; ramas florales primarias 12–14, 7.0–8.1 × 0.6–0.8 cm, a 40–45° de inclinación; 2.0–3.1 × 0.3–0.5 cm en ramas secundarias; bractéolas 3.2–4.1 mm; pedicelos 1.0–2.0 mm. **Flores** amarillas; ovario 5.0–6.1 × 1.9–2.4 mm de largo, cilíndrico, con cuello 0.1–0.2 × 0.1–0.2 mm; tubo 0.5–0.8 mm de profundo, cilíndrico, más ancho en la mitad, ligeramente acanalado; pedicelos 0.5–0.6 × 0.2–0.3; brácteas 0.2–0.4; tépalos 2.1–2.6 × 0.3–0.6 cm, lanceolados, redondeados en el ápice, los exteriores más largos, sobrepuestos en los interiores, persisten flexionados y coriáceos cuando secos, los interiores con una quilla interior; filamentos exertos 1.5 veces más largos que los tépalos; anteras tan largas como los tépalos, de color amarillo, el ápice de color oscuro. **Infrutescencias** hasta de 3.4–5.2 m, con ramas compactas y cortas 3.7–6.9 × 0.5–1 cm, en ángulo de 55–65° con respecto al raquis, en sentido del reloj. **Cápsulas** maduras, 2.3–2.5 × 1.5–1.9 cm, subglobosas anchamente elipsoides, duras, de coloración sepia rojizo, con cierto tomento gris en la base. **Semillas** 5.0–6.0 × 3.0–5.0 mm, en forma de letra D, ligeramente acanalada en el centro, de color negro.

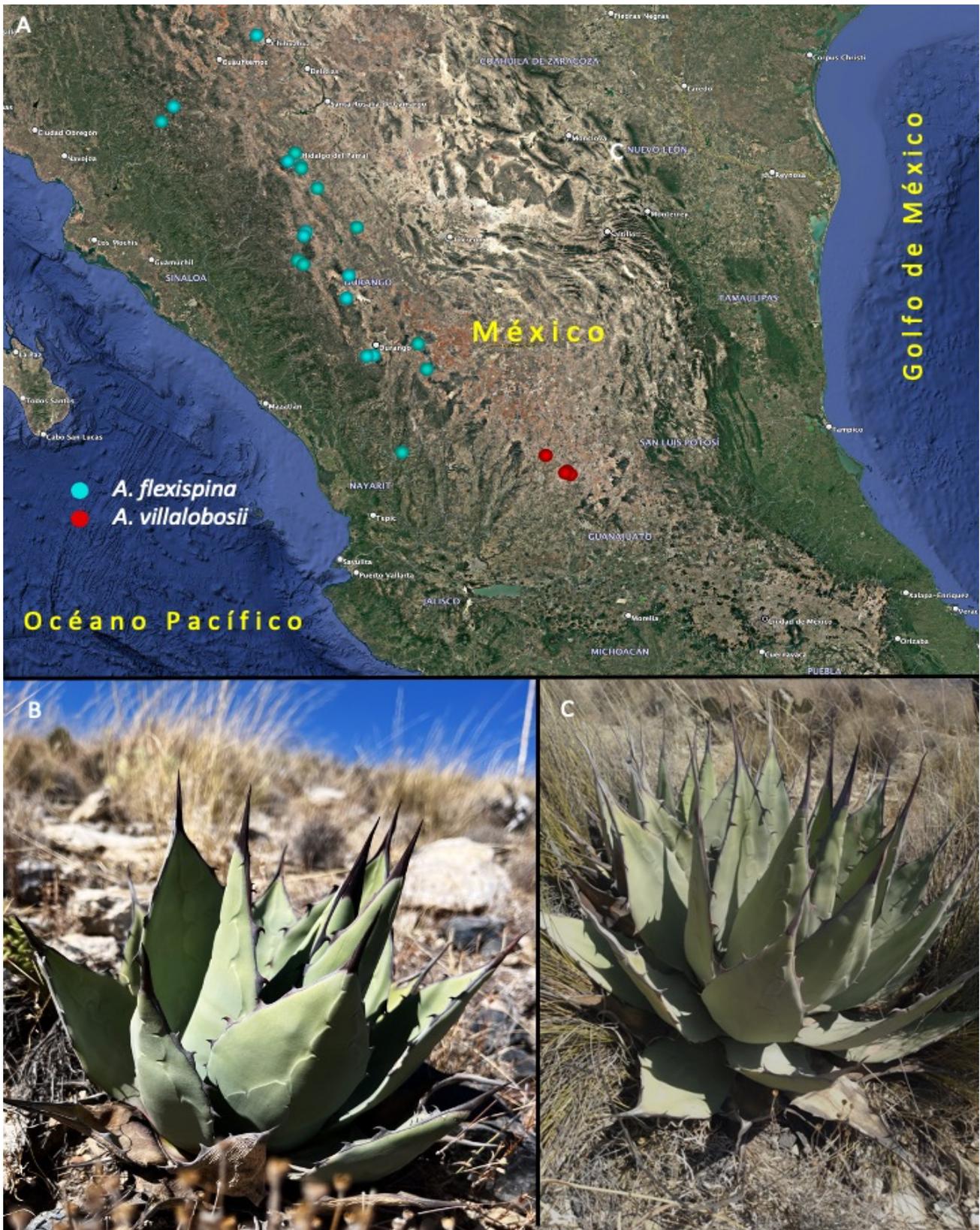

**Figura 1**. A. Distribución geográfica de *Agave flexispina* y *A. villalobosii.* B–C. Rosetas de *A. villalobosii.* Fotografías B–C, por F. Trigo Raya.

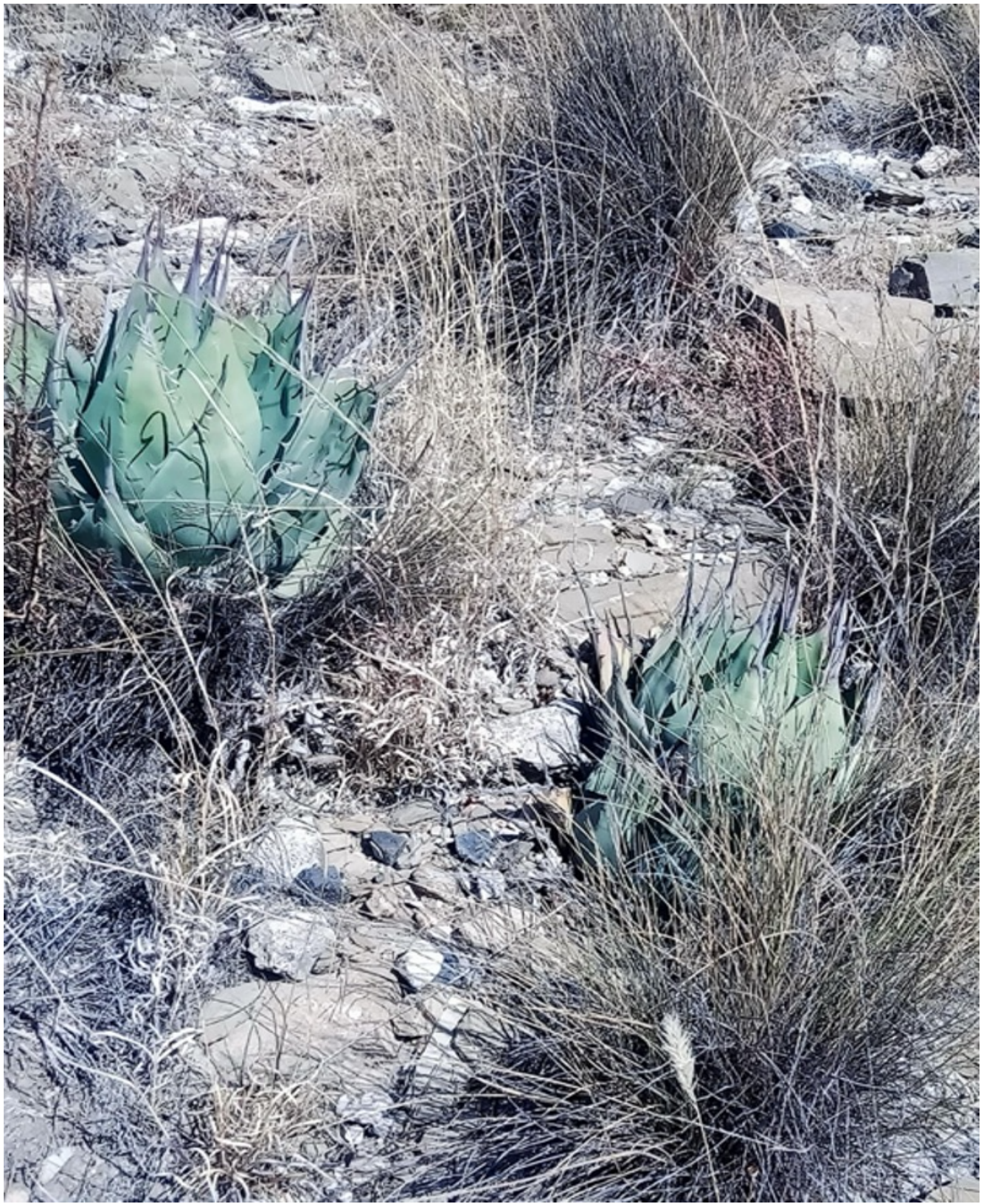

**Figura 2**. *Agave flexispina* in situ, Tepezalá, Rincón de Romos, Aguascalientes. Fotografía de P. Oropeza Gutiérrez

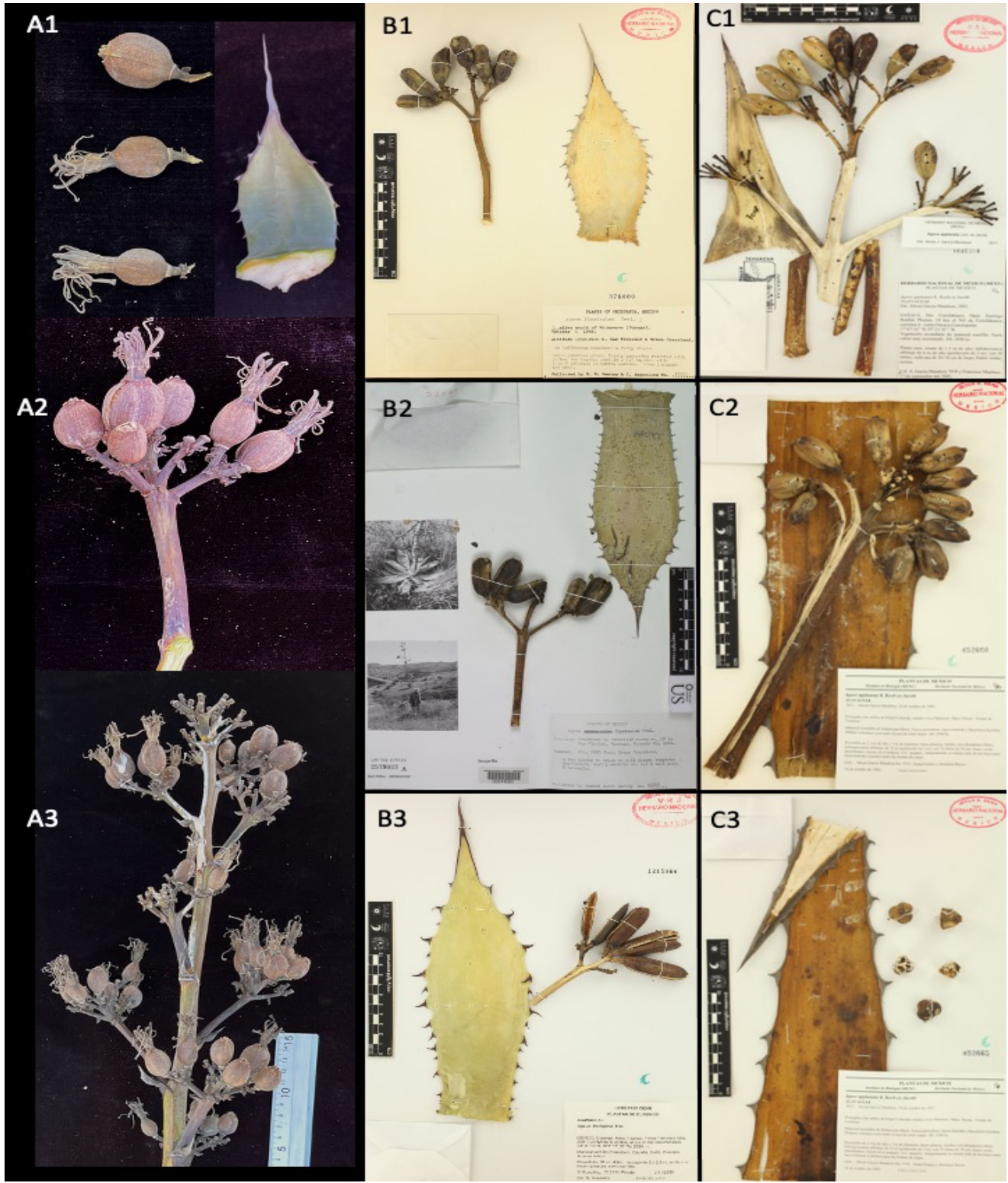

**Figura 3**. A1–A3. *Agave villalobosii*, de Tepezalá, Rincón de Romos, Aguascalientes, de material tipo. Fotografías: A1–A3 de J. A. Vázquez García. B. *A. flexispina*: B1, de Matamoros, Chihuahua, *Gentry & Arguelles 17922* (MEXU); B2, de Canutillo, Durango, *Gentry 22049* (US); B3, de Poanas, Durango, *González 7131* (MEXU). C1–C3. *A. applanata*: C1, de Coixtlahuaca, Santiago Ihuitlán Plumas, Oaxaca, *García Mendoza 7018* (MEXU); C2–3, de Perote, Veracruz, *García Mendoza 5741* (MEXU).

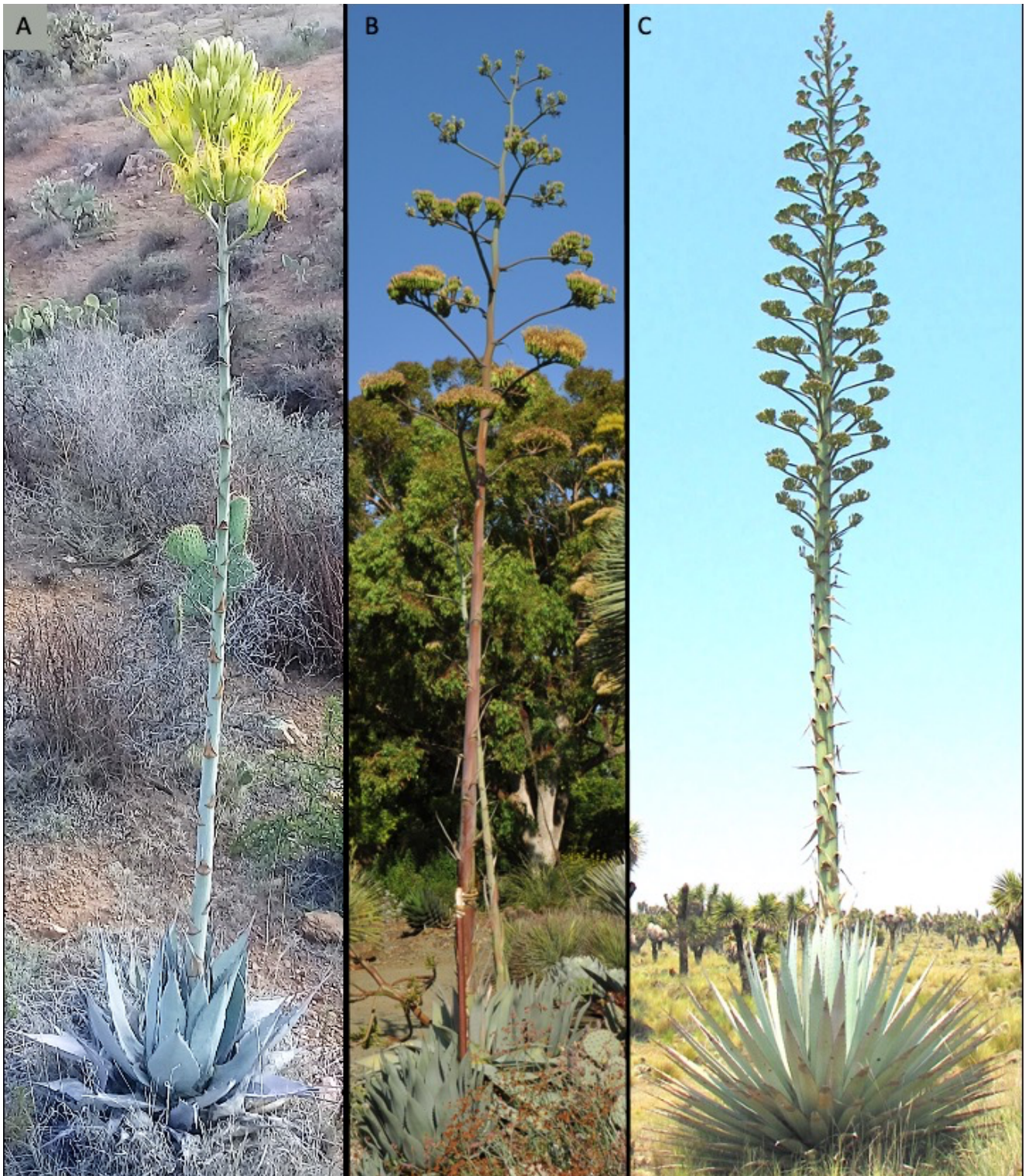

**Figura 4.** Inflorescencias: A. *Agave villalobosii*, Guadalupe, Zacatecas, México. https://mexico.inaturalist.org./observations/39556046, Fotografía A: de Quirino, 29 de junio de 2019. B. *Agave flexispina*, en el Ruth Bancroft Garden and Nursery (RBG&N), CA. Fotografía B: de B. Kemble, Curador en el RBG&N, https://www.ruthbancroftgarden.org/plants/agave-flexispina/. C. *Agave applanata*, https://mexico.inaturalist.org/observations/309390804, El Frijol Colorado, Perote, Ver., México, Fotografía C: de A. González Gallina, 4 de junio de 2010.

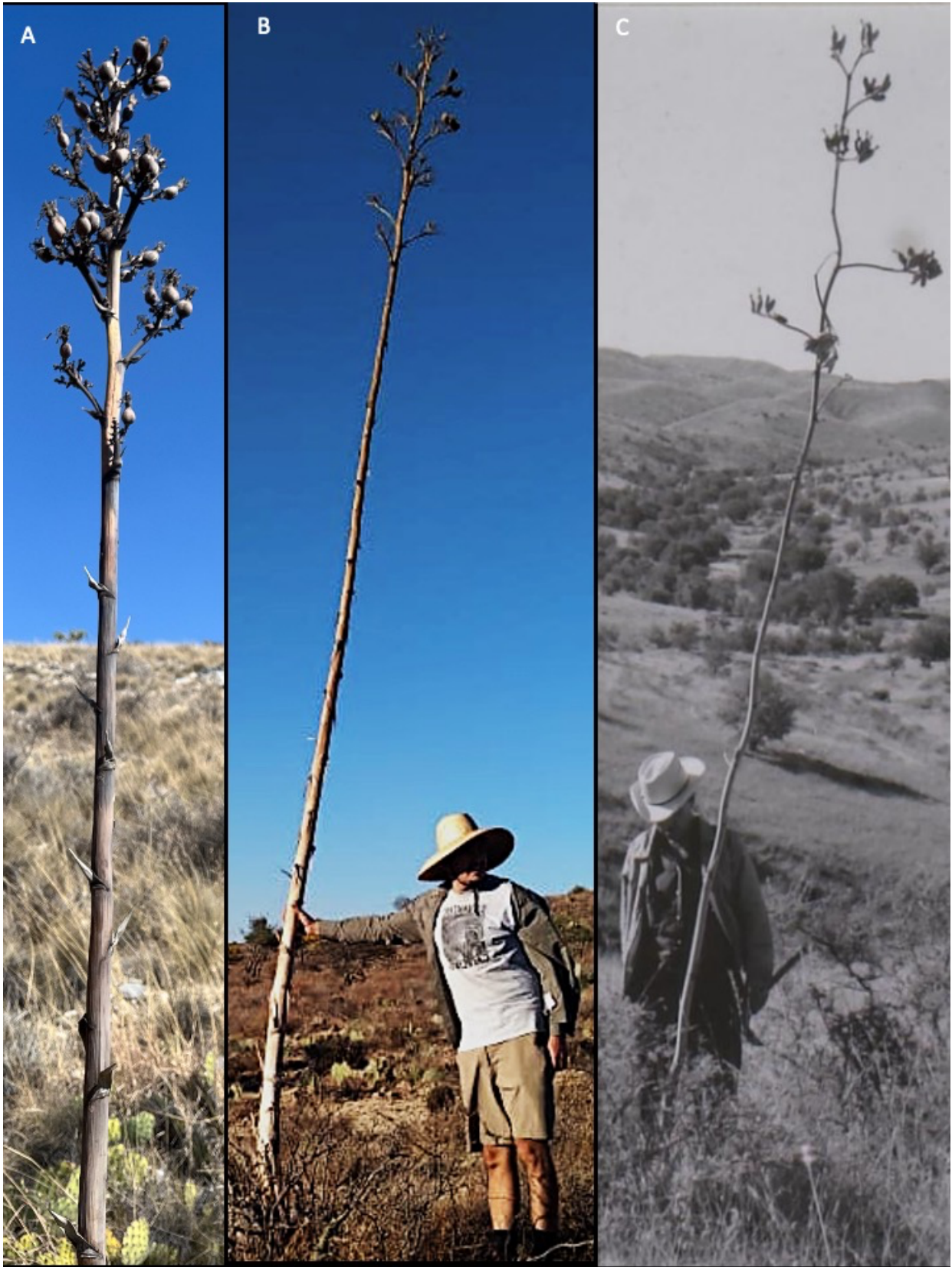

**Figura 5:** Infrutescencias. *Agave villalobosii,* de Tepezalá (A) y Tepetates (B), Aguascalientes. C. *A. flexispina,* de Canutillo, Durango, *Gentry 22049* (US). Fotografías: A de F. Trigo Raya y B. de J. L. Luquín Arce.

**Cuadro 1**. Diferencias entre *Agave villalobosii* y especies morfológicamente relacionadas.

| | *Agave villalobosii* | *A. flexispina* | *A. applanata* |
|---|---|---|---|
| **Rosetas** | | | |
| **No. de hojas por roseta** | 25–30 | 35–40 | 60–66 |
| Altura de roseta (dm) | 3.0–4.0 | 2.5–3.5 (-45) | 5.0–1.2 |
| Diámetro de roseta (dm) | 5.0–7.0 | 50.0–70.0 | 10.0–20.0 |
| Color | Verde grisaceo | Glauco a amarillo verdoso | Gris glauco |
| **Hojas** | | | |
| Forma | Obovadas a elípticas | Ovadas, acuminadas | Lanceoladas a lineales |
| Márgenes | semi-rectos e incurvados | Ondulados a crenados | Córneos o ausentes |
| Largo de dientes (mm) | 4.0–8.0 | 5.0–8.0 | 8.0–15.0 |
| Tipo de diente | Afilados, curvos, retrorsos, pruinoso grisáceos, delgados | Mamilados y retrorsos marrón a pruinoso | Fuertes y afilados, casi rectos marrón oscuro, pruinosos cerosos |
| Separación de dientes (mm) | 2.0–2.2 | (0.7–)11.0–15.0 | 4.0–6.0 |
| Espina /color | Acanalada en la parte baja, dura, ondulada, grisácea | Flexuosa, plana, acanalada / marrón a gris pruinoso | Fuerte, plana, ahuecada encima, marrón rojizo oscuro |
| Largo de espina largo (cm) | 3.4–6.1 | 2.5–3.5 | 3.0–7.0 |
| **Inflorescencia** | | | |
| Longitud de panícula (m) | 1.50–1.65 | 2.5–3.5(–6) | 3.5–12.0 |
| Color de flor | Amarillas | Amarillo verdoso, ápice de palos exteriores con tinte rojizo | Amarillas con ovario verdoso |
| Largo del ovario (mm) | ¿? | 22.0–35.0 | 35.0–38.0 |
| Longitud del tubo (mm) | 5.0–8.0 | 13.0–18.0 | 15.0–22.0 |
| Longitud de tépalos (mm) | 21.0–26.0 | 10.0–18.0 | 15.0–22.0 |
| Longitud de filamentos (mm) | 31.0–0.36 | 40.0–50.0 | 45.0–55.0 |
| Longitud de anteras (mm) | ¿? | 17.0–23.0 | 23.0–30.0 |
| **Infrutescencia** | | | |
| Longitud (m) | 3.4–5.2 | 5.0–7.0 | 6.0–14.0 |
| No. de ramas primarias (umbelas) | 10–12 | 6–12(–18) | 55–60 |
| Inclinación de ramas sobre eje x (°) | 40–45 | 20-35 | 45–60 |
| Ramas primarias desarrolladas (cm) | 3.7–6.9 × 0.5–1.1 | 9.2–10.6 × 0.8–0.9 | 26.0–33.0 × 2.0–3.6 |
| Tamaño de cápsulas (cm) | 2.3–2.5 × 1.5–1.9 | 3.5–4.5 × 1.5–1.7 | 4.0–4.8 × 2.2–2.4 |
| Tamaño de semillas (mm) | 5.0–6.0 × 3.0–5.0. | 5.5–7.0 × 4.0–5.0 | 7.7–7.9 × 5.0–6.0 |
| **Distribución** | | | |
| Estados / elevación (m) /floración | Ags. y oriente de Zac./ 2100–2350 / jun.–jul. | Chih. Dgo. Zac. Nay. / 1300–2440 / jun.–sep. | Chih. Dgo. Oeste de Zac. Gto. Qro. Hgo. Méx., Pue., Ver. Oax. / 2300 / jun.–oct. |

**Eponimia:** La especie es dedicada a Iván Villalobos-Juárez, su descubridor, quien ya sospechaba que se trataba de una especie nueva y nos guió a dos de sus poblaciones de Aguascalientes. Iván es un gran explorador y curador de colecciones vivas en el Jardín Comunitario de El Valle de Las Delicias, Rincón de Romos, Aguascalientes, donde alberga una importante colección botánica de los mezcales de México. Además, ha encabezado la lucha para la conservación del hábitat de la ranita de madriguera (*Smilisca dentata*) y la concientización de las comunidades del centro de Aguascalientes sobre el cuidado y preservación de reptiles, anfibios y peces nativos. Además, es fundador de la organización "Los Hijos del Desierto", que busca rescatar la bioculturalidad tradicional de Aguascalientes con la elaboración de platillos y recetas que han sido olvidadas por las comunidades del Estado.

***Distribución y ecología***: Crece en la Meseta central mexicana en los estados de Zacatecas (Guadalupe) y Aguascalientes (Cerro Capulín, Tepezalá, Rincón de Romos; y El Tepetate, San Francisco De Los Romo), en pastizales o matorral xerófilo de clima semiárido y templado, con *Agave applanata, A. angustifolia* (*"samandoki"*), *A. filifera*, *A. salmiana*, *Tymophylla setifolia, Russelia* sp., *Stenocereus* sp. y *Stenocactus* aff. *coptonogonus*, *Mammillaria* sp., *Ferocactus histrix*, *Echinocereus pectinatus*, *Opuntia robusta*, *Yucca filifera* y *Cylindropuntia tunicata*.

**Estado de conservación**: Utilizando un ancho de celda de 2 km en GeoCAT (Bachman *et al.* 2011), *Agave villalobosii* tiene una Extensión de Ocurrencia (EOO) de 495,579 km² y un Área de Ocupación (AOO) de 12.000 km². Según las categorías y criterios de la UICN (2025), la especie se considera en Peligro Crítico debido a su pequeña EOO. *A. villalobosii* es una especie muy escasa, aún en su localidad tipo, donde las rosetas se encuentran dispersas y solitarias. Además, con base al conteo de semillas de tres cápsulas, la mayoría de estas resultaron estériles (solo el 3% fértiles), quizá por la baja polinización o autoincompatibilidad, lo cual pone en mayor riesgo a la especie. Algunas semillas se han sembrado en el invernadero-vivero de plantas nativas amenazadas, de la Universidad de Guadalajara-CUCBA, Zapopan.

**Especímenes y observaciones examinadas.** MÉXICO: Aguascalientes: El Tepetate, San Francisco De Los Romo, 8 Marzo 2025 (fr.), *J. Antonio Vázquez García 10405, con I. Villalobos Juárez, P. J. Oropeza Gutiérrez, J. L. Luquin Arce y F, Trigo Raya* (IBUG). Zacatecas, Guadalupe, 29 junio 2019 (fl.), fotografía de *Quirino* (https://mexico.inaturalist.org./observations/39556046).

## DISCUSIÓN

Las diferencias de las poblaciones con respecto a especies morfológicamente similares y geográficamente cercanas, *Agave flexispina*, y *A. applanata*, tanto vegetativas como reproductivas (Cuadro 1, Fig. 1), son evidentes y suficientes para respaldarlas y reconocerlas como una especie nueva. Además, se registra formalmente por primera vez la ocurrencia de *A. applanata* (*Vázquez-García et al. 10406*, IBUG) en el estado de Aguascalientes, el cual no había sido citado ni en los trabajos sobre el género *Agave* para Aguascalientes y occidente de México (Gentry 1982; De La Cerda-Lemus 2004, 2007; Vázquez-García *et al.* 2007b), ni en el inventario de la flora de México (Villaseñor 2016). Lo anterior evidencía que la exploración botánica y los estudios taxonómicos son aún muy necesarios en la meseta central mexicana. *A. villalobosii* se diferencía fácilmente de *A. applanata* (Cuadro 1) por tener menor número de hojas por roseta (25–30 vs. 60–66), rosetas de menor tamaño (3.4–4.0 × 5.0–7.0 vs. 5.5.0–12.0 × 10.0–20.0 dm); hojas obovadas a elípticas vs. lanceoladas a lineales; dientes más cortos (0.4–0.8 vs. 8.0–15.0 mm); menor espaciamiento entre dientes (2.0–2.2 vs. 4.0–6.0 cm); tubo floral más corto (5.0–8.0 vs. 15.0–22.0 cm); filamentos más cortos (31.0–36.0 vs. 45.0–55.0 mm); panícula más corta (1.4–1.6 vs. 4.0–8.0(–14.0) cm); menor número de ramas primarias (10–12 vs. 55–60); ramas primarias de menor tamaño (3.7–6.9 × 0.5–1.1 vs. 26.0 × 33.0 cm); cápsulas más cortas (2.3–2.5 vs. 4.0–4.8 cm) y menor tamaño de semillas (5.0–6.0 × 3.0–5.0 vs. 7.7–7.9 × 6.0 mm).

Aunque no se observaron flores en fresco, de la nueva especie, las mediciones de flores fueron hechas en los restos florales aún unidos al fruto, los cuales fueron rehidratados previamente para su estudio. Es de resaltar que aunque las inflorescencias observadas son de 1.50 a 1.65 m (en junio, apenas establecidas las lluvias), la panícula continúa creciendo durante la fructificación, pudiendo alcanzar hasta 5.2 m (a finales del verano e inicio de otoño), presuntamente debido al sostenido abastecimiento hídrico durante el verano.

La distribución geográfica de la especie, por ahora confinada a Aguascalientes y oriente de Zacatecas (Fig. 1), es preliminar. Con base en observaciones recientes de imágenes de rosetas juveniles en la plataforma inaturalista.org, se espera que la especie pueda ser confirmada en los estados de Jalisco, Guanajuato y otros estados del centro de México. En caso de que su distribución resulte más extensa, como resultado de nuevas exploraciones, su diagnóstico preliminar de estado de conservación en peligro crítico también cambiaría. Una expedición a Ojuelos de Jalisco, realizada en octubre de 2025, en busca de poblaciones potenciales de *A. villalobosii*, no resultó exitosa, ya que se trataba de poblaciones juveniles de *A. parryi*. Sin embargo, la búsqueda debe continuar en otros sitios de Ojuelos y Lagos de Moreno, en Jalisco, así como en el estado de Guanajuato.

Este trabajo resalta la importancia de fomentar el establecimiento de jardines comunitarios bioculturales, donde se concentran especies nativas locales, lo cual en este estudio permitió avanzar la ciencia, al conocer la inflorescencia enana (1.5 m) de la especie transplantada de Tepezalá, Rincón de Romos, la cual tenía 7 años en cultivo (desde 2011), generándose así la sospecha de Iván Villalobos Juárez de que podría tratarse de una especie no descrita, hipótesis ahora bien sustentada con esta investigación.